\documentclass[a4paper,fleqn,usenatbib]{mnras}

\usepackage[T1]{fontenc}
\usepackage{ae,aecompl}

\usepackage{graphicx}	
\usepackage{amsmath}	
\usepackage{amssymb}	

\newcommand{\degree}{^{\circ}}

\title[Methanol and water masers from IRAS 19410+2336]{Methanol and water maser observations separate disc and outflow sources in IRAS 19410+2336}

\author[M. Darwish et al.]{
Darwish, M. S.$^{1,2}$,\thanks{E-mail: darwish.msk@gmail.com}
Edris, K. A.$^{3}$,
Richards, A. M. S.$^{4}$,
Etoka, S.$^{4}$,
\newauthor
Saad, M. S.$^{1,2}$,
Beheary, M. M.$^{3}$,
Fuller, G. A.$^{4}$
\\
$^{1}$Astronomy Department, National Research Institute of Astronomy and Geophysics (NRIAG), 11421 Helwan, Cairo, Egypt.\\
$^{2}$Kottamia Center of Scientific Excellence in Astronomy and Space Science (KCScE, STDF No. 5217, ASRT), Cairo, Egypt.\\
$^{3}$Astronomy and Meteorology Department, Faculty of Science, Al-Azhar University, Cairo, Egypt.\\
$^{4}$Jodrell Bank Centre for Astrophysics, Department of Physics $\&$ Astronomy, The University of Manchester, M13 9PL, UK.\\
}
\date{Accepted XXX. Received YYY; in original form ZZZ}

\pubyear{2019}

\begin{document}
\label{firstpage}
\pagerange{\pageref{firstpage}--\pageref{lastpage}}
\maketitle
%
\begin{abstract}
We investigate the kinematics of high mass protostellar objects within the high mass star forming region IRAS 19410+2336.
{We performed high angular resolution observations of 6.7-GHz methanol and 22 GHz water masers using the MERLIN (Multi-Element Radio Linked Interferometer Network) and e-MERLIN interferometers.}
{The 6.7-GHz methanol maser emission line was detected within the $\sim$ 16--27 km s$^{-1}$ velocity range with a peak flux density $\sim$50 Jy. The maser spots are spread over $\sim$1.3 arcsec on the sky,  corresponding to $\sim$2800 au at a distance of 2.16 kpc. These are the first astrometric measurements at 6.7 GHz in IRAS 19410+2336. The 22-GHz water maser line was imaged in 2005 and 2019 (the latter with good astrometry). Its velocities range from 13 to $\sim$29 km s$^{-1}$. The peak flux density was found to be 18.7 Jy and 13.487 Jy in 2005, and 2019, respectively. The distribution of the water maser components is up to 165 mas, $\sim$350 au at 2.16 kpc.}
{We find that the Eastern methanol masers most probably trace outflows from the region of millimetre source mm1. The water masers to the West lie in a disc (flared or interacting with outflow/infall) around another more evolved millimetre source (13-s). The maser distribution suggests that the disc lies at an angle of 60$^{\circ}$ or more to the plane of the sky  and the observed line of sight velocities then suggest an enclosed mass between 44 M$_{\odot}$ and as little as 11 M$_{\odot}$ if the disc is edge-on. The Western methanol masers may be infalling.}
\end{abstract}


%
\section{Introduction}\label{sec:intro}

The ambiguity of how massive stars were formed makes studying their formation one of the most critical current issues in astrophysics.
Both observation and theory face many challenges \citep{2007prpl.conf..165B,2007ARA&A..45..481Z}. The challenge from the observational side is due to several reasons: Firstly, they form at  large distances ($>1$ kpc, with the exceptions of the Orion \citep{2007PASJ...59..897H} and Cepheus A \citep{2012arXiv1204.3808T} regions). Secondly their formation happens on much shorter time scales than their low mass counterparts. Furthermore, high-mass star formation takes place inside high density clouds, obscuring our view of the process.
Theoretically, the fundamental problem in high-mass star formation is known as the "radiation pressure problem". That is, the radiation pressure from the forming protostellar object is predicted to stop the accretion process  when the star reaches a limit $\sim$10M$\odot$ \citep{1987ApJ...319..850W}. Since more massive stars do exist, the theoretical models must deal with this challenge to explain how the mass build up proceeds.
\par So far, the most widely accepted model for high-mass star formation is based on a 'scaled-up' version of the low-mass model, but with a few alterations \citep[e.g.][and references their in]{2010MNRAS.406..102K}. This model suggested that in order for stars to gain enough mass to be considered high-mass stars, they must undergo much higher accretion rates through a disc which allows the radiation pressure to be overcome \citep{2004ASPC..322..263T,2009Sci...323..754K}.
\par Interstellar masers are believed to be one of the best tracers of massive star forming regions  \citep{2007IAUS..242..213E} and the study of many maser transitions has shed a lot of light on the high-mass star formation (HMSF) process.
Masers are used to characterize circumstellar discs, bipolar outflows \citep{2000prpl.conf..327S}, magnetic fields \citep[e.g.][]{1984Obs...104..125C,1990ApJ...357..502M, 2005A&A...434..213E,2012MNRAS.423..647E} and internal motions \citep[e.g.][]{2009ApJ...705.1548R,2010A&A...517A..71S, 2011PASJ...63...53S}. \citet{2007IAUS..242..213E} provide a rudimentary time line for classifying the evolutionary stage of young high-mass stars.
\par H$_2$O water and CH$_3$OH class II methanol masers are  believed to be closely associated with the early stages of massive star formation \citep{1997MNRAS.289..203C,2004A&A...414..235S}. It is known for quite some time now that class-II 6.7-GHz methanol masers are commonly found associated with ground state OH masers within $\simeq$1 arcsec radius \citep{1996MNRAS.279...79C}. High angular resolution imaging has revealed that this association can reach $\sim$100 mas radius but close association (i.e., down to a $\simeq$10 mas radius) is rare \citep{2005MNRAS.360.1162E} due to the combination of two factors: the important role played by grain-mantle evaporation in the production of both OH and CH$_3$OH during HMSF \citep{1995MNRAS.272..184H} and the expected evolutionary offset in the appearance and duration of these two types of maser emission, where the results of large surveys \citep{2010MNRAS.401.2219B} and \citep{2010MNRAS.406.1487B} suggest that although OH and methanol can be found in the same regions in agreement with the models of \citet{2002MNRAS.331..521C}, methanol masers appear first and disappear while OH and water are still detectable.
\par Another well known association is that of water and Class-II 6.7-GHz methanol, with their exclusive association (i.e., no sign of OH maser emission) thought to be tracing earlier stages of the star formation process \citep{2007IAUS..242..213E,2010MNRAS.401.2219B}. The spatial correlation between H$_2$O and CH$_3$OH masers towards a number of sources has been studied by \citet{2002A&A...390..289B}, who concluded that although the typical separation  between H$_2$O and CH$_3$OH masers is 0.03 pc there is no evidence for closer associations. They felt that this was unlikely, given that the maser species are collisionally and radiatively pumped, respectively.
\citet{2018MNRAS.474.3898B} compared the few-hundred sources in the HOPS (H$_2$O Southern Galactic Plane) and MMB (Methanol Multibeam) surveys and found that a majority of associated methanol and water masers occurred within 1\farcs0, or 0.024 pc (at a typical distance of 5 kpc), comparable to the combined astrometric accuracy.

Additionally, these masers can be detected originating from deep inside  the dusty molecular envelope of high-mass stars in early stages of formation, making them  a sensitive probe for discovering stars in their embryonic state.
\par Considering these issues, we present high angular resolution observations of H$_2$O and CH$_3$OH masers toward the high mass star forming region IRAS 19410 +2336. The previous observations carried out toward the IRAS 19410+2336 region suggest that it is a young massive star forming region in an early evolutionary state with a complex morphology \citep[and references therein]{2008ApJ...685.1005Q}.
\citet{1994AAS...184.3013K} reported that there is no Ultracompact H~{\small II} region detected at 8.4 GHz, while \citet{2002ApJ...566..931S} detected  weak 3.6 cm continuum emission of 1 mJy.

\citet{2002ApJ...566..945B}, using the IRAM 30-m MAMBO bolometer (pixel spacing $\sim5\farcs5$), found a $\lambda$ 1.2-mm source mm1 at $\alpha$(J2000) = 19$^h$ 43$^m$ 11$^s$.24, $\delta$(J2000) = 23$\degree$ 44$^\prime$ $3\farcs4$. The total astrometric accuracy of the peak is $0\farcs5$ or better but the source is extended over $>5''$. One of the $\lambda$ 3-mm sources mapped by  \citet{2012A&A...545A..51R} using the IRAM interferometer has a similar extent and position,
$\alpha$(J2000) = 19$^h$ 43$^m$ 11$^s$.206, $\delta$(J2000) = 23$\degree$ 44$^\prime$ $3\farcs29$. A number of their detections at $\lambda$ 1.2 mm fall in this region, the closest being 13-s at $\alpha$(J2000) = 19$^h$ 43$^m$ 11$^s$.206, $\delta$(J2000) = 23$\degree$ 44$^\prime$ $3\farcs29$, with an astrometric error estimated at $0\farcs18$.

\par The 6.7-GHz methanol maser toward IRAS 19410+2336 was first discovered by \citet{1991ApJ...380L..75M}. He found 6.7-GHz emission in the velocity range from 15 to 28 km s$^{-1}$ with a peak flux density of 103 Jy.  \citet{1993MNRAS.262..343M} reported a similar detection and also monitored the 12-GHz transition, which had a peak flux density of 10.6 Jy at 27 km s$^{-1}$. The first recorded imaging of this region at 22 GHz, in 1979, was by \citet{1981ApJ...243..769L}. They reported that the water maser had a velocity range from 12 to 20.6 km s$^{-1}$ with a peak flux density of 21 Jy.
\par The distance to this source is estimated to be 2.16 kpc \citep{2009A&A...507.1117X}, while the total infrared luminosity is around 10$^4$ ${L_\odot}$ \citep{2008A&A...489..229M}.
The average temperature for the whole region estimated to be 40$\pm$15 K, while  the temperature of the dense cores is 80$\pm$40 K \citep{2012A&A...545A..51R}.
\par To investigate what the CH$_3$OH  and H$_2$O masers trace and how they are distributed and correlated with other tracers, IRAS 19410+2336 has been observed using MERLIN and e-MERLIN at high angular resolution. The details of the observations and data reduction are presented in Sec. 2 while the results are introduced in Sec.3. The discussion is given in Sec. 4, and conclusions drawn in Sec. 5.

\begin{table*}
\caption{MERLIN and e-MERLIN observational parameters for methanol and water masers.}
\label{tab:MRLIN_data}
\begin{center}
\renewcommand{\arraystretch}{1.1}
\begin{tabular}{|c|c|c||c|}
\hline
 Observational parameters   & CH$_3$OH  &\multicolumn{2}{|c|}{H$_2$O}\\
\hline
Date of observation & 31-Dec-2004 & 01-Jan-2005 &  20-March-2019\\
No. antenna & Five antennas & Five antennas & Five antennas\\
Field centre ($\alpha$, $\delta$ J2000)&$\!\!$19$^h\!$ 43$^m\!$ 11$^s$.247, 23$\degree\!$ 44$^\prime\!$ $03\farcs32$$\!\!$ & \multicolumn{2}{|c|}{$\!\!$19$^h\!$ 43$^m\!$ 11$^s$.184, 23$\degree\!$ 44$^\prime\!$ $02\farcs961$$\!\!$}\\
Rest frequency (MHz)& 6668.518 & 22235.079 & 22235.079 \\
No. of frequency channels & 511& 255 & 512 \\
Total band width (MHz) & 0.5 & 4 & 4\\
Bandpass calibrator & 0552+398 & 3C273 & 3C84 \\
Phase calibrator & 1932+204A & 1932+204A & J1946+2300 \\
Restoring beam (mas$\times$mas)& 70$\times$45 & 20$\times$20 & 29$\times$28 \\
$\!\!\!\!$rms in quiet channel (Jy beam$^{-1}$)$\!\!\!\!$ & 0.0346 & 0.0707     & 0.050 \\
\hline
\end{tabular}
      \end{center}
      \end{table*} %

\section{Observations and data reduction}\label{sec:obs}
Our initial observations of  IRAS 19410+2336 were  carried out using the MERLIN interferometer. Observations in dual-polarization (LL,RR) of the Class-II 6.7-GHz methanol and 22-GHz water maser emission were performed in December 2004 and January 2005, respectively. In both cases, the target data were adjusted to fixed velocity with respect to the LSR (Local Standard of Rest) in the correlator. The data were extracted from the MERLIN archive and converted to FITS at Jodrell Bank Center for Astrophysics (JBCA) using local software (dprogs) and the AIPS (Astronomical Image processing system) software package. CASA (the Common Astronomy Software Application) software package version 5.1.1 was used in order to calibrate and reduce the data. The observational parameters for IRAS 19410+2336 are listed in Table \ref{tab:MRLIN_data}. The observations and data reduction followed normal MERLIN procedures, see the MERLIN User Guide \citep{Diamond03}.
\par Subsequent observations of the 22-GHz water maser emission were performed with eMERLIN in March 2019; the data were reduced manually in CASA following a strategy similar to that described in
\emph{https://github.com/e-merlin/eMERLIN\_CASA\_pipeline}.

\subsection{CH$_3$OH masers}
The 6.7-GHz methanol line was observed in a spectral bandwidth of 0.5 MHz corresponding to 30 km s$^{-1}$ velocity range with a channel separation of 0.04 km s$^{-1}$. The source 0552+398 was observed as a bandpass calibrator. At this frequency the flux density for 0552+398 was set to be 4.6 Jy based on previous scaling using 3C286, \citep{1977A&A....61...99B}, allowing for the resolution of MERLIN.
\par As a phase reference calibrator, the compact Quasar 1932+204A was observed through the observations at
$\alpha$(J2000) = 19$^h$ 35$^m$ 10$^s$.473, $\delta$(J2000) = 20$\degree$ 31$^\prime$ $54\farcs154$ with a cycle of phase reference:target 1.7:5.3 min, in a useful bandwidth of 13 MHz overlapping the line band.
\par 0552+398 was observed in both bandwidths and used to measure the phase offset between configurations. This correction, along with the bandpass table and the phase reference solutions for phase and amplitude, were applied to the target. The brightest target channel, at $V_{\mathrm{LSR}}$ 26.9 km s$^{-1}$ was used for self-calibration and the solutions were applied to all channels. The absolute position errors for 6.7-GHz methanol masers are 12 mas in each of RA and Dec, derived from the error contributions described by \citet{Diamond03}.
\par The position used for astrometry was measured before self-calibration but any changes were negligible.
The CLEAN algorithm as implemented in the CASA package "tclean" was used for cleaning and de-convolution of the image cube. This was exported as FITS to allow component fitting in AIPS. We edited the velocity labeling to correct inconsistencies between CASA and AIPS in order to ensure that the data for analysis were correctly labeled in velocities with respect to the LSR. Some parameters of the final cube are given in Table \ref{tab:MRLIN_data}.
Table \ref{tab:Positions} lists the absolute positions as well as the corresponding velocities of the brightest feature for both methanol and water masers.
\par In order to fit the components through the different channels in a cube image we used the AIPS task SAD (Search And Destroy). The positions of the maser components were determined via 2D Gaussian fitting above 3$\sigma$ threshold in each channel map. The components were required to  appear in at least three consecutive channels with positions within 20 mas of their occurrence in each consecutive channel, such groups forming  spectral features. The peak intensities of these features with their relative positions and velocities were consider to represent the group. {The spectra in Figs. 1 and 2 show the component flux densities for each separate group. In most cases the spectral profiles are approximately Gaussian although some are blended.}

\subsection{H$_2$O masers}
The 22 GHz water maser line was observed in 2005 with a spectral bandwidth of 4 MHz corresponding to 50 km s$^{-1}$ velocity range with a channel separation of 0.21 km s$^{-1}$.
\par In order to calibrate the variation of instrumental gain and phase across the spectral bandpass, the source 3C273 was selected to be a bandpass and flux calibrator.
The flux density for 3C273 was set to be 24 Jy (flux monitoring kindly supplied by Tersranta, Metsahovi, private communication).
Observations of 1932+204A were again carried out for phase referencing but unfortunately it was too faint and far away to be useful at 22 GHz.
The brightest maser channel was therefore used for self-calibration and the solutions applied to all channels, but this means that there is no improvement of the astrometry over the input observing position.
\par A short e-MERLIN observation was made in 2019 using a spectral bandwidth of 4 MHz in 512 channels, giving a channel separation of 0.105 km s$^{-1}$, at constant frequency. The upgraded array allows greater sensitivity due to new receiver systems, a broader continuum bandwidth thanks to optical fibers and a new, more flexible correlator. Thus, we could also observe calibration sources at 125 MHz bandwidth. 3C84 was used as a bandpass and flux scale calibrator (assuming a flux density of 20 Jy) and also to
calculate the phase offset between the 4- and 125-MHz configurations. J1925+2106, 4.8$^{\circ}$ from the target, flux density about 2 Jy beam$^{-1}$, was observed every 20 min as time-dependent amplitude calibrator. J1946+2300 was used as the phase reference source, at a position of
$\alpha$(J2000) = 19$^h$ 46$^m$ 06$^s$.2510, $\delta$(J2000) = 23$\degree$ 00$^\prime$ $04\farcs414$ (a separation of
 0.99$^{\circ}$ from the target) with a flux density of 0.07 Jy beam$^{-1}$.
\par We applied the corrections derived from these calibrators to IRAS 19410+2336. The IRAS 19410+2336 dataset was then transformed to constant velocity in the LSR frame. The best few hours of data were used to image a channel at 23.7
km s$^{-1}$ which contained a bright peak at $\alpha$(2000)= 19$^h$ 43$^m$ 11$^s$.184, $\delta$(2000)= 23$\degree$ 44$^\prime$ $03\farcs008$, at a signal-to-noise ratio of 23, giving a stochastic position error of 2 mas.
\par The largest quantifiable contribution to position uncertainty is the phase change in the offset between phase-reference and target, contributing 4 mas. Allowing for other errors, due to antenna position uncertainties and calibration errors, the astrometric accuracy of this position is better than 10 mas. This peak was then used for self-calibration, applying the corrections to all line data, and a data cube made with a restoring beam of
29 mas$\times$28 mas, rms $\sim50$ mJy except in the brightest channels
which were dynamic range limited.
\par Other aspects of data reduction were similar to that for methanol and ``tclean" was used to make a final data cube; some parameters are given in Table~\ref{tab:MRLIN_data}.
We fitted components to the emission and grouped the spots into features as described in Section 2.1, with a positional match requirement of $\sim$20 mas.
\begin{table}
\caption{The absolute positions and the radial velocities of the brightest maser feature at epochs 2005 (methanol) and 2019 (water). The positions leading terms are $\alpha$(J2000) = 19$^h$ 43$^m$, $\delta$(J2000) = 23$\degree$ 44$^\prime$.}\quad
\label{tab:Positions}
\begin{center}
\renewcommand{\arraystretch}{1.1}
\begin{tabular}{|c|c|c|c|}
\hline
 Type of maser   &  $\alpha$(J2000) & $\delta$(J2000) & $V_{\mathrm{LSR}}$ (km s$^{-1}$)\\
\hline
CH$_3$OH & 11.238$\pm$0$^s$.013 & 3.297$\pm$0$^{\prime\prime}$.015 & 26.548 \\
H$_2$O  & 11.1840$\pm$0$^s$.1 & 2.961$\pm$0$^{\prime\prime}$.1 & 26.420   \\
\hline
\end{tabular}
     \end{center}
        \end{table} %

\section{Results}
The main aim of this work was to understand  better the kinematics of the high mass star forming region IRAS 19410+2336 as well as to investigate the possibility of determining the position of the embedded protostellar objects.
 The 6.7-GHz CH$_3$OH and 22 GHz water masers were detected using MERLIN and e-MERLIN.
 In addition to the distribution of each maser type, the positional and velocity relationships between the two masers species were examined. Further details about each maser type are discussed below.

\begin{figure}
	\centering
  		\includegraphics[width = .95 \linewidth] {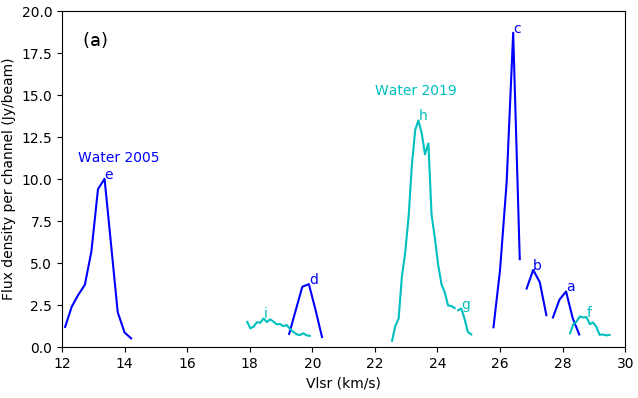}  		\includegraphics[width = .95\linewidth]{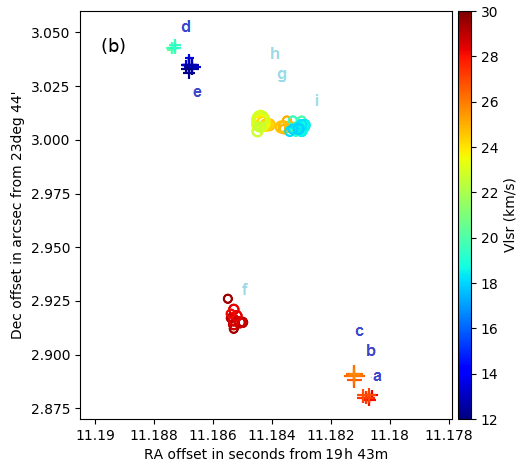}
   \caption{\label{fig:spec}Panel (a) shows the spectra of H$_2$O maser associated with IRAS 19410+2336, the numbers represent the features listed in table 3, the blue spectra is for the 2005 and the cyan one represents the 2019 epoch, while (b) represents the distribution of the H$_2$O maser components toward the IRAS 19410+2336 region for 2005 epoch (plus signs) and 2019 epoch (circles), the size of the symbol is proportional to the log of their intensity.}
        \end{figure}

\subsection{Water masers}

The H$_2$O masers in the IRAS 19410+2336 region were observed at two epochs, 2005 and 2019. Five spectral features were identified in 2005 while four features were found in 2019.
Table~\ref{tab:properties} lists the peak intensities, positions and velocities for both water maser epochs.
Figure~\ref{fig:spec} (a) represents the spectral components observed in both epochs. In 2005, the velocity range covered by the emission was 12--29 km s$^{-1}$, {grouped into 5 features} with the highest flux density determined to be 18.7 Jy and the faintest feature peak 3.304 Jy.
 \par In 2019 we identified four features with peaks in the range 18.44 -- 28.76 km s$^{-1}$, with peak flux densities ranging from 1.6 to 13.5 Jy. {Features 2005 a and d have counterparts close in velocity in 2019 f and i, at around  19 and 28 km s$^{-1}$. 2005 features b and c lie between a and f but are slightly less red-shifted whilst  2019 features g and h lie between d and i and are slightly less blue-shifted. The most blue-shifted feature, e, was only detected by us in 2005.}
\par Figure~\ref{fig:spec} (b) shows that the water maser components are distributed within an area about 165 mas across, $\sim$350 au at 2.16 kpc.
The masers detected in 2019 are about 90 mas (200 au) in angular extent, also with the more blue-shifted emission to the North. The mid point of the extreme velocity peaks is $\alpha$(J2000) = 19$^h$ 43$^m$ 11$^s$.1840, $\delta$(J2000) = 23$\degree$ 44$^\prime$ $2\farcs961$.
\par At all epochs, the 22 GHz H$_{2}$O maser peaks fall within a velocity
range between 12--29 km s$^{-1}$ with a maximum angular extent of 200 mas. The more blue-shifted features are always the most northerly. Only the 2019 data have astrometrically accurate positions. We assumed that the masers are centred on a common centre of excitation, at an intermediate position and velocity. We aligned the 2005 data
with the 2019 data by assuming that the red- and blue-shifted clumps were offset symmetrically about a common centre at both epochs.
%
%
\begin{table*}
\caption{The properties of the detected H$_2$O  maser features toward IRAS 19410+2336 with relative position errors, the positions leading terms are $\alpha$(J2000)  = 19$^h$ 43$^m$, $\delta$(J2000) = +23$\degree$ 44$^\prime$. The top and bottom halves are 2005 and 2019 data, respectively.}\quad
\label{tab:properties}
\begin{center}
\renewcommand{\arraystretch}{1.1}
\begin{tabular}{|c|c|c|c|c|c|c|c|}
\hline
Epoch & No.  &  Vel.(km s$^{-1}$) & Flux(Jy) & $\alpha$ (s)& Error(ms)& $\delta$ ($''$)& Error(mas)\\
\hline
2005 &a   &28.11 &   3.304   $\pm$   0.070 &     11.1807   	& 0.016   & 2.879  &   0.236\\
&b   &    27.06 &    4.588   $\pm$   0.079 &     11.1807  	& 0.012   & 2.881  &   0.188 \\
&c   &    26.42 &    18.708  $\pm$   0.078 &     11.1812 	& 0.002   & 2.890  &   0.045  \\
&d   &    19.89 &    3.743   $\pm$   0.079 &     11.1873  	& 0.014   & 3.043  &   0.227  \\
&e   &    13.36 &    10.013  $\pm$   0.119 &     11.1868  	& 0.008   & 3.035  &   0.129  \\
\hline
2019 & f& 28.76 &    1.784   $\pm$   0.102 &     11.1851  & 0.036   & 2.915  &      0.5\\
&g   &    24.76 &    2.287   $\pm$   0.162 &     11.1836  & 0.094   & 3.006  &     0.7  \\
&h   &    23.39 &    13.487  $\pm$   0.205 &     11.1844  & 0.007   & 3.008  &     0.1  \\
&i   &    18.44 &    1.699   $\pm$   0.135 &     11.1829  & 0.072   & 3.007  &     1.0   \\
\hline
\end{tabular}
     \end{center}
        \end{table*}
The results, especially the disappearance of the 13 km s$^{-1}$
feature in 2019, confirm the suggestion by
\citet{1995MNRAS.272...96C,2000A&AS..143..269S} that the region is strongly
variable. We discuss in Section 4 whether the other
features at similar velocities can be related between epochs.

\subsection{Methanol masers}
Our 6.7-GHz masers observations represent the first astrometric measurements, since the positions quoted by \citet{2002A&A...390..289B} are derived from \citet{2000A&A...362.1093M} and are actually for the 12 GHz masers.
We detected 20 6.7-GHz methanol maser features over a velocity range
of 16.3 -- 27.6 km s$^{-1}$, with a complex spatial distribution.
Figure~\ref{fig:Meth_spectra} (upper) illustrates the spectra of the 6.7-GHz masers, where the brightest feature (1) $\sim$50 Jy has the maximum velocity $\sim$27 km s$^{-1}$, while the faintest peak is $\sim$0.388 Jy at velocity 20.46 km s$^{-1}$. {Fig.~\ref{fig:Meth_spectra} (lower) }
 shows that our 6.7-GHz methanol masers are extended over $\sim1\farcs3$, with the greatest elongation E-W. The brightest emission, features 1 and 18, are separated by $\sim$$0.''8$ at a position angle of
$\sim$60$^{\circ}$ (from N through E), with the most red-shifted emission
to the East. The weaker, mostly blue-shifted methanol emission covers
a range of position angles extending to $\sim$110$^{\circ}$. These
orientations are close to the position angles of the outflows
identified as South C and A by \citet{2003A&A...408..601B}.
\par Weaker emission at intermediate velocities is more scattered,
mostly but not exclusively associated with the locations of the peaks. The components within a 50 mas region around $\alpha$(J2000) = 19$^h$ 43$^m$ 11$^s$.25, $\delta$(J2000) = 23$\degree$ 44$^\prime$ $03\farcs3$
cover the greatest localized methanol maser velocity span, of 19--27.5 km s$^{-1}$.

\par The parameters of the 6.7 GHz CH$_3$OH maser spectral features are listed in Table~\ref{tab:properties_Meth},
while Figure~\ref{fig:Meth_distribution}  (a and b) represents the distribution of the methanol and water masers components throughout the region, color-coded by velocity.
\begin{table*}
\footnotesize
\caption{The properties of the detected CH$_3$OH  maser features toward IRAS 19410+2336 with position errors. The position leading terms are $\alpha$(J2000)  = 19$^h$ 43$^m$, $\delta$(J2000) =  +23$\degree$ 44$^\prime$.}\quad
\label{tab:properties_Meth}
\begin{center}
\renewcommand{\arraystretch}{1.1}
\begin{tabular}{|c|c|c|c|c|c|c|c|c|}
\hline
 No.  &  Vel.(km s$^{-1}$) & Flux(Jy) & $\alpha$ (s)& Error(ms)& $\delta$ ($''$)& Error(mas) \\
\hline
1  &  26.92   &   49.183  $\pm$ 0.079  &   11.2464 & 0.007  &    3.298 &  0.1   \\
2  &  24.46   &   10.165  $\pm$ 0.070  &   11.2500 & 0.007  &    3.288 &  0.1   \\
3  &  23.67   &   5.763   $\pm$ 0.066  &   11.2500 & 0.014  &    3.296 &  0.2   \\
4  &  21.65   &   0.514   $\pm$ 0.077  &   11.2524 & 0.211  &    3.113 &  3.0   \\
5  &  21.43   &   3.257   $\pm$ 0.066  &   11.1958 & 0.021  &    2.916 &  0.3   \\
6  &  20.60   &   1.221   $\pm$ 0.078  &   11.1958 & 0.087  &    2.934 &  1.3   \\
7  &  20.07   &   9.683   $\pm$ 0.075  &   11.1956 & 0.007  &    2.942 &  0.1   \\
8  &  19.67   &   19.902  $\pm$ 0.084  &   11.1950 & 0.007  &    2.951 &  0.1   \\
9  &  21.65   &   0.496   $\pm$ 0.090  &   11.2533 & 0.335  &    3.312 &  4.1   \\
10 &  20.99   &   0.494   $\pm$ 0.103  &   11.2716 & 0.364  &    3.263 &  6.7   \\
11 &  20.46   &   0.388   $\pm$ 0.060  &   11.2519 & 0.160  &    3.328 &  1.5   \\
12 &  20.42   &   0.555   $\pm$ 0.077  &   11.2514 & 0.211  &    3.283 &  2.9   \\
13 &  20.07   &   2.880   $\pm$ 0.078  &   11.2744 & 0.029  &    3.398 &  0.4   \\
14 &  19.72   &   2.949   $\pm$ 0.087  &   11.1874 & 0.036  &    3.138 &  0.5   \\
15 &  19.23   &   1.033   $\pm$ 0.122  &   11.2477 & 0.233  &    3.294 &  4.0   \\
16 &  18.88   &   4.842   $\pm$ 0.069  &   11.2714 & 0.014  &    3.243 &  0.2   \\
17 &  18.97   &   9.960   $\pm$ 0.076  &   11.2688 & 0.007  &    3.218 &  0.1   \\
18 &  16.95   &   30.107  $\pm$ 0.106  &   11.1960 & 0.007  &    2.943 &  0.1   \\
19 &  16.82   &   9.814   $\pm$ 0.148  &   11.2672 & 0.021  &    3.250 &  0.5   \\
20 &  17.08   &   19.015  $\pm$ 0.111  &   11.2673 & 0.007  &    3.208 &  0.1   \\
\hline
\end{tabular}
     \end{center}
        \end{table*}

\begin{figure}
  \includegraphics[angle = 0,width = 9cm]{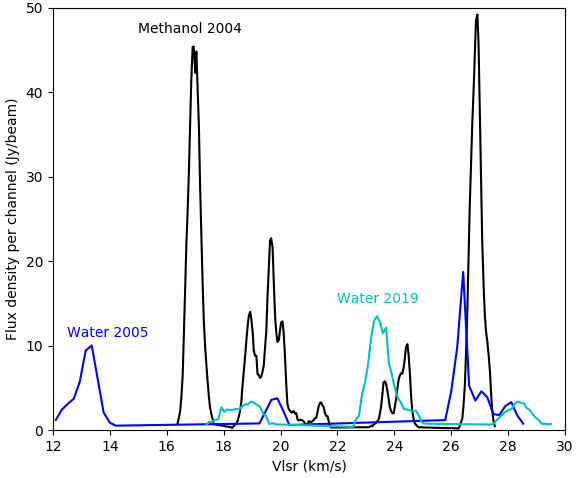}
  \caption{\label{fig:spec_MethWat}The integrated spectra of the 6.7 GHz CH$_3$OH and 22 GHz water masers for both epochs 2005 and 2019}.
     \end{figure}

\begin{figure}
  \centering
  \includegraphics[width = .95\linewidth]{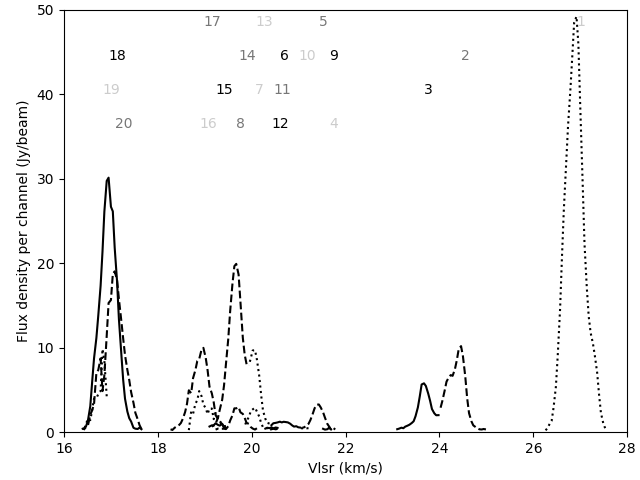}
  \includegraphics[width = .95\linewidth]{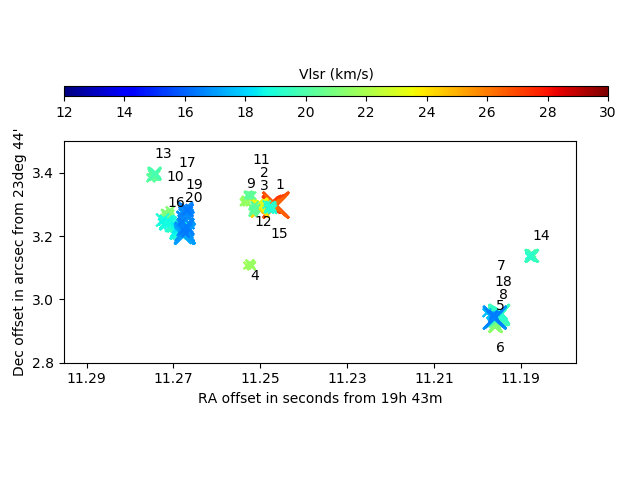}
  \caption{\label{fig:Meth_spectra}The upper panel represents the spectra of 6.7-GHz methanol maser emission, the numbers refer to the features listed in Table~\ref{tab:properties_Meth}. Numbers in black, dark and mid grey correspond to full, dotted and dashed lines. The lower panel shows the distribution of the CH$_3$OH maser components toward the IRAS 19410+2336 region, the size of the symbol is proportional to the log of their intensities.}
\end{figure} %
It also  shows that the methanol maser spots are spread over $\sim1\farcs3$ corresponding to $\sim$2808 au at a distance 2.16 kpc. The radial velocities were found to be in the range of 16.38 km s$^{-1}$ to 27.53 km s$^{-1}$.
\par The spatial correlation between water and methanol is closest for the West part of the blue-shifted methanol masers, see Figure~\ref{fig:Meth_distribution}(b), {but the closest angular separation is $>0\farcs1$. The red-shifted methanol maser components, Figure~\ref{fig:Meth_distribution}(a),  are  offset by  $>0\farcs5$ from the water masers, with more blue-shifted methanol still further East. The flux densities for the methanol peaks range from 0.514 Jy to 49.183 Jy, where the brightest maser is the most red-shifted, located in between the more blue-shifted clumps.}

\begin{figure*}
\includegraphics[width = 0.5\linewidth]{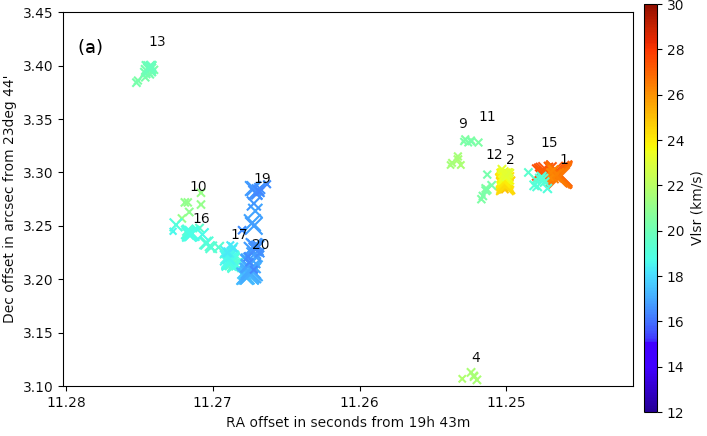}
\includegraphics[width = 0.48\linewidth]{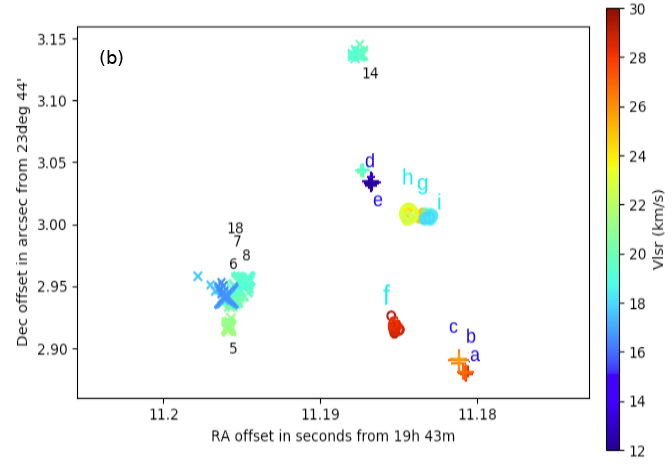}
    \caption{The distribution of the 6.7 GHz CH$_3$OH(crosses) and H$_2$O maser components denoted by plus signs and circles for the 2005 and 2019 epochs respectively  toward the IRAS 19410+2336 region. The size of the symbol is proportional to the log of the maser component intensity, colour-coded by the LSR velocities. Panels (a) and (b) represent the Eastern and Western regions where masers were imaged, respectively.}
    \label{fig:Meth_distribution}
\end{figure*}

\section{Discussion}
The IRAS 19410+2336 was previously observed by \citet{1995MNRAS.272...96C} where 6.7 GHz and 12 GHz masers were detected in velocity ranges 14 to 27 km s$^{-1}$ and 16 to 28 km s$^{-1}$  respectively, while the peak flux densities were found to be 42 Jy and 15.8 Jy, respectively. \citet{2000A&AS..143..269S}, using single dish observations, found that the strongest feature of 34 Jy occurred at 17.3 km s$^{-1}$ and the spectrum had changed considerably compared with the previous observations (see Figure 1 in \citealt{2000A&AS..143..269S}).
\par Figure \ref{fig:spec_MethWat} shows the integrated spectra of 6.7-GHz methanol and 22-GHz water masers. The water masers appears to cover a wider range in LSR velocity than the methanol but the methanol peaks are brighter (during our observations). Figure \ref{fig:all_features}, right, plots our maser observations, denoting methanol by crosses and water masers by plus signs (2005) and circles (2019).  The size of methanol and water feature symbols is proportional to the log of their intensities, color-coded by velocity.  Methanol masers at 12 GHz  are shown by triangles  \citep {2000A&A...362.1093M}. The c3mm-sII and 13-s mm continuum peaks are shown by  solid stars (position accuracies $0\farcs5, 0\farcs2$) \citep{2012A&A...545A..51R}. The mm1 1.2-mm continuum peak is shown by an unfilled star, (position accuracy 0\farcs5) \citet{2003A&A...408..601B}. The peak of the thermal SiO clump is shown by a square (measured in a beam size of 4'') \citep{2016A&A...589A..29W}.
This shows that the 12-GHz masers observed by \citet{2000A&A...362.1093M}, have an angular separation of $\sim$ $0\farcs7$ and appear to be spatially correlated (within astrometric uncertainties) with the brightest 6.7 GHz peaks at similar velocities. It is likely that they are tracing the same phenomena around the embedded protostellar object/s (see later).
\par Turning to the 22 GHz water masers, Figure~\ref{fig:spec} (b)
shows that the 2019 components have a smaller separation and are
rotated with respect to the 2005 data.  The displacement of each
feature is about 70 mas or 150 au, which would imply a proper motion
speed of 36 km s$^{-1}$ if the clumps are physically the same at both
epochs. This seems unlikely for two reasons. Firstly, although it
could signify infall, or rotation of a NE-SW disc, these mechanisms
would have to be very non-radial or warped. Secondly, the speed is
far greater than the $V_{\mathrm{lsr}}$ span, and  would imply an unreasonably large enclosed mass, see discussion in  Section~\ref{sec:discandinfall} para 4.
\par
\begin{figure*}
   \includegraphics[angle = 0,width = 12 cm]{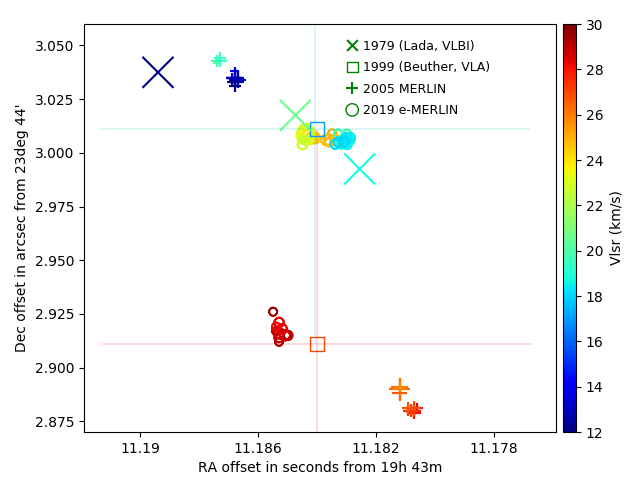}
   \caption{\label{fig:water_all} Comparison between previous and
     current water maser imaging towards IRAS 19410+2336. The X are for \citet{1981ApJ...243..769L} VLBI measurements, the squares are for \citet{2002A&A...390..289B} measurements,
while the crosses (+) and the circles are for our MERLIN and eMERLIN measurements respectively.
The epochs are aligned as outlined in Sections 3.1 and 4.}
        \end{figure*}

 {\citet{2002A&A...390..289B} used 1999 VLA observations
to measure the positions of the H$_{2}$O maser peaks.  The peaks at
$V_{\mathrm{LSR}}$ 27 and 17 km s$^{-1}$ were located at
$\alpha$(J2000)  = 19$^h$ 43$^m$ 11.2$^s$, $\delta$(J2000) =  +23$\degree$
44$^\prime$ $03\farcs0$ and $\alpha$(J2000)  = 19$^h$ 43$^m$ 11.2$^s$,
$\delta$(J2000) =  +23$\degree$ 44$^\prime$ $03\farcs1$, respectively, with
an absolute astrometric error estimated at $1''$ and a relative
accuracy of $0\farcs1$.}  This gives a separation of (0, 100) $\pm$ 100
mas, with the more blue-shifted peak to the north, as in our
data. {The mid point of these peaks is at (11.2$^s$, $03\farcs05$)
in the coordinates of Figure \ref{fig:water_all}. This is offset by (220, 44)
mas from our 2019 mid point at (11$^s$.1840 $2\farcs961$), which is well
within the astrometric uncertainty of the VLA data.} It seems
reasonable to align the mid-point of \citet{2002A&A...390..289B}'s
data with our observations, as in Figures \ref{fig:water_all} and
\ref{fig:all_features}.
\par The 22 GHz maser toward IRAS 19410+2336 was first resolved by
\citet{1981ApJ...243..769L}, as previously mentioned. Their position
(in J2000) is over 10 arcsec from  recent measurements, probably
due to the difficulties in determing astrometric positions from fringe-rate analysis in
1981. The maser velocities cover a range overlapping our 2005
observations, at 12.0, 18.7 and 20.6 km s$^{-1}$. The 12 km s$^{-1}$
feature was brightest, at $\sim$21 Jy, the others, of a few Jy, being
offset from this by (--94, --40) and (--64, --20) mas, uncertainties $\le$10
mas. We aligned the average position of these features with the
average position of the features we detected at similar velocities in
2019.
\par Comparing the observations of \citet{1981ApJ...243..769L} and \citet{2002A&A...390..289B} with our data shows that emission around 18--20 km s$^{-1}$ was detected at all epochs. The most blue-shifted feature, around 13 km s$^{-1}$, was detected in 1979 and 2005 but not in 1999 nor 2019, and the red-shifted features were detected in 1999, 2005 and 2019.
We note that 22 GHz single-dish observations by \citet{2002ApJ...566..931S} found a peak of 110 Jy, and single-dish observations by \citet{2015MNRAS.453.4203X} detected emission between 3--33 km s$^{-1}$ with a peak $\sim$28 Jy but in both cases it is possible that emission from other parts of the region is in the beam. If the features at similar velocities are from the same gas at different epochs (which seems unlikely, see Section 4.2), their flux densities vary roughly 2-fold between epochs; if they are from different parts of IRAS 19410+2336 then its water maser variability is even more pronounced. The flux variability could be caused by the variability of infrared sources associated with this region.



\par Our 2005 and 2019 astrometric observations, with accuracies of 12 and 10 mas respectively, locate most of the 22-GHz H$_2$O masers firmly on the Western edge of the 6.7 and 12-GHz methanol maser distribution.
\par In conclusion we can claim that for \citet{1981ApJ...243..769L} observations the astrometry was uncertain, being based on only a single baseline, so it is possible that the emission detected by \citet{1981ApJ...243..769L} and by \citet{2002A&A...390..289B} as well as ours came from a similar region. The flux variability could be caused by variable infrared source that associated with this region.

\par Class I methanol masers at 95 and 44 GHz  have been detected towards this region by \citet{1995ARep...39...18V} and \citet{2011ARep...55.1086L}. The peak flux density at 95 GHz was estimated to be 3.9 Jy at a  velocity of 22.59 km s$^{-1}$, while \citet{2011ARep...55.1086L} reported three maser features with the brightest feature at 17 Jy  at a peak velocity of 23.6 km s$^{-1}$.
The presence of Class I methanol masers associated with shocked material in protostellar outflows (see \citealt{2003A&A...408..601B}) in addition to Class II methanol masers is a further evidence for tracing very early stages of massive star formation \citep{2009ApJ...702.1615C} and \citet[and references therein]{2016A&A...592A..31L}.

\par The OH maser was first reported by \citet{1987A&A...181...19B} with flux density 1.1 Jy and corresponding radial velocity 20.9 km s$^{-1}$.  \cite{2007A&A...465..865E}, using the Greenbank single dish telescope, detected peaks at flux densities of 1.7 Jy and 0.7 Jy in velocity ranges of 19 to 22 km s$^{-1}$ and 20 to 23 km s$^{-1}$ respectively, showing modest variability.
The position was $\alpha$(J2000) = 19$^h$ 43$^m$ 12$^s$.2 $\pm1^s.1$, $\delta$(J2000) = 23$\degree$ 44$^\prime$ $3''$ $\pm12''$
{As noted in Section 1, the presence of OH masers implies a later evolutionary stage than does methanol class II and H$_2$O without OH.  This suggests that different prototars in this region are at different evolutionary stages, but high-resolution OH imaging is needed to confirm which masers are associated with which objects.}

\par Figure \ref{fig:all_features} (right) shows the positions of other emission detected in this region.The thermal SiO \citep{2016A&A...589A..29W} was observed with a 4-arcsec beam, larger than the total maser extent. Millimetre source mm1, position accuracy $0.''5$ \citep{2003A&A...408..601B} is located close to the center of the methanol maser distribution whilst 13-s, position accuracy $0.''13$ \citep{2012A&A...545A..51R} is close to the water masers.{
We suggest that the Eastern methanol masers (Figure 3 (a)) are associated with thermal SiO and a less evolved protostar than the Western methanol and water masers.  This is explained in the following Sections 4.1 and
4.2.}
%

\subsection {Outflow activity}
\label{sec:outflow}
Outflow activity in high-mass star forming regions is predicted by different massive stars formation scenarios.
The competitive accretion model suggests that outflows in  high mass star forming regions should be  less collimated  than those produced by low mass star forming regions \citep{2014prpl.conf..149T}.
\par On the other hand, the core accretion model, which is considered as a scaled-up version of low-mass star formation, suggests well-collimated outflows similar to those associated with low mass protostars \citep[and references therein]{2016A&A...589A..29W}. According to \citet{2007prpl.conf..245A} the opening angles for well-collimated outflows in massive star forming regions are between 25$\degree$ and 30$\degree$.
\par The outflow activity in IRAS 19410+2336 has been studied extensively by \citet{2003A&A...408..601B,2008A&A...489..229M} and \citet{2016A&A...589A..29W}.
\citet{2003A&A...408..601B} observed CO (J = 1-0)
in addition to the 2.6 mm continuum
using the IRAM 30 m telescope and the Plateau de Bure interferometer,
and the H$_2$ line at 2.12 $\mu$m, using the Calar Alto
3.5 m telescope.
 These results support the presence of
well-collimated outflows similar to those found in low-mass star
formation. They concluded that the region contains a strong outflow
activity with at least seven outflows. They also report that the
kinematics of the individual outflows resemble those in low mass-star
forming regions.

\par The line widths of the SiO  detected by \citet{2016A&A...589A..29W} in the region (typically 18 km s$^{-1}$) along with a Gaussian profile, but with an additional signal in the line wings, strongly suggests that SiO traces dynamical gas in jets and/or outflows \citep{2016A&A...589A..29W}.

\par
\begin{figure*}
    \begin{center}
    \includegraphics[angle = 0,width = 7cm,clip]{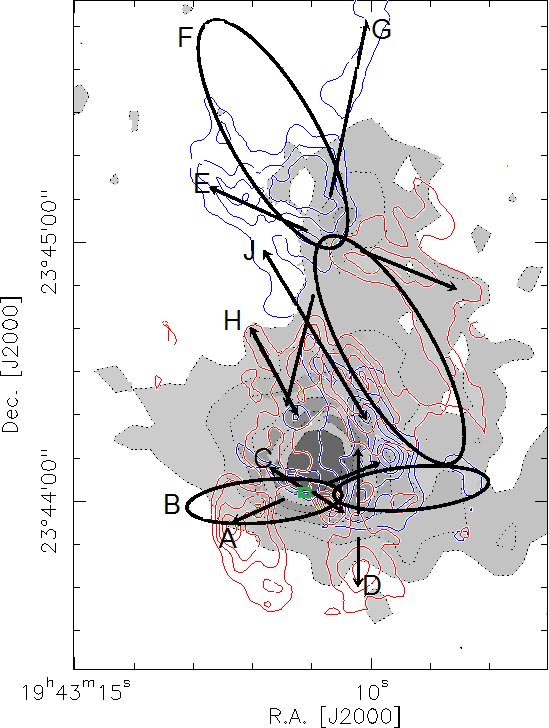}
    \includegraphics[angle = 0,width = 10.5cm,clip]{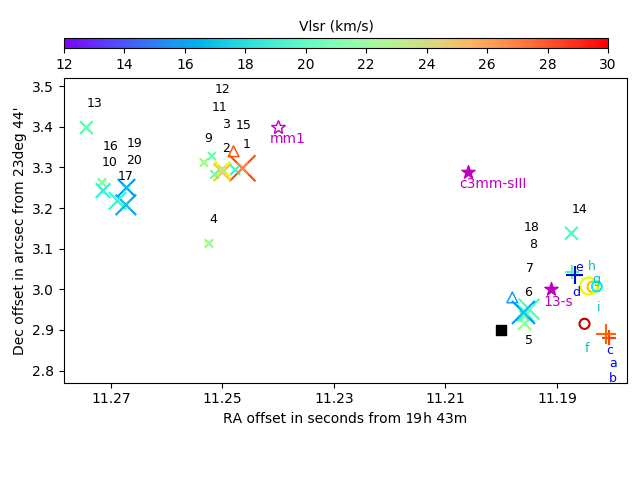}
    	\end{center}
    		\caption{\label{fig:all_features}  Right hand side: Species in IRAS 19410+2336 detected close to the masers imaged by (e-)MERLIN. Methanol masers (crosses) and water masers (plus signs and circles) (this work). The size of methanol and water feature symbols is proportional to the log of their intensities, color-coded by velocity. 12 GHz methanol masers (triangles) \citep {2000A&A...362.1093M}.  c3mm-sII and 13-s mm continuum peaks (solid stars, position accuracies $0\farcs5, 0\farcs2$) \citep{2012A&A...545A..51R}. mm1 1.2-mm continuum peak (unfilled star, position accuracy 0\farcs5) \citep{2003A&A...408..601B}. Peak of the $\sim4''$ thermal SiO clump (square) \citep{2016A&A...589A..29W}. The left hand side represents CO (1-0) data from \citet{2003A&A...408..601B} which appears in the background of the map while the ellipsis and arrows represent the proposed outflow directions from \citet{2003A&A...408..601B}. The maser area is marked by the green square.}
      		  \end{figure*}

 The methanol masers show the most red-shifted emission near the center of the distribution with more blue-shifted emission extending to the SE and SW.
The peak position of SiO, shown in Fig.~\ref{fig:all_features} was measured with a $4''$ beam, is thought to trace an outflow.  Its location points to a close association with the Eastern methanol maser features.
It is therefore likely that these maser features probe the same outflow as the SiO. The mid-point of the red- and blue-shifted methanol masers is centred at about
$\alpha$(J2000)  = 19$^h$ 43$^m$ 11.26$^s$, $\delta$(J2000) =  +23$\degree$ 44$^\prime$ $03\farcs3$). This is within the  $0\farcs5$ position uncertainty of the mm1 peak  \citep{2002ApJ...566..945B}.
 \citet{2003A&A...408..601B} suggest that mm1 produces multiple outflows. The maser distribution could trace ongoing outflow activity, in a similar direction to H$_2$ feature (2), associated with outflow C by \citet{2003A&A...408..601B} (Figure \ref{fig:all_features}). The distribution of the masers with respect to their mid-point suggests outflow opening angles of about 40$^{\circ}$ and 20$^{\circ}$. This results agrees with the well-collimated outflow criteria in  \citet{2007prpl.conf..245A}.
\par \citet{2015MNRAS.449..119D} presented an evolutionary sequence based on the association of 6.7-GHz methanol maser with  outflows in the high-mass star forming region. This sequence suggested that:
Firstly, the outflow appears and grows without any presence of methanol maser.
Then, both outflow and methanol masers are present and could be observed.
Later, methanol maser emission disappears but the outflow remains detectable and the UC~H~{\small II} region appears.
Eventually, the outflow switches off while the UC~H~{\small II} region remains.

Although \citet{2015MNRAS.449..119D} didn't test the relation between water masers and class II methanol masers in the presence of outflows, \citet{2007IAUS..242..213E} concluded that methanol class II masers in high mass star forming regions are tracing an earlier evolutionary phase than water masers.

\par Bearing in mind the scheme of \citet{2015MNRAS.449..119D} and also  the  maser based evolutionary sequence of \citet{2007IAUS..242..213E}, we suggest that the embedded protostar associated with the Eastern methanol masers (Figure \ref{fig:all_features})  is in an early stage of evolution. These masers lie within the fitted extent of the 3-mm source detected by \citet{2012A&A...545A..51R}. The closest 1.2-mm source to the Eastern masers is the peak of mm1 detected by \citet{2002ApJ...566..945B}. We suggest that the Western methanol and water masers are associated with a more evolved 1.2-mm source, 13-s, which is located close to the centre of the water maser distribution  (within the combined position errors of $0\farcs2$) (Section 4.2).  \citet{2012A&A...545A..51R} estimated the temperature of 13-s at 90 K, the hottest source in the region at about twice the temperature assumed by \citet{2003A&A...408..601B} for mm1.  13-s is also associated with a higher gas column density at 25 10$^{23}$cm$^{-2}$, at the location where \citet{2008A&A...489..229M} and \citet{2008ApJ...685.1005Q} detected a near-infrared(NIR) and mid-infrared(MIR) source.

mm1 and 13-s are separated by $0\farcs8$, greater than their combined, nominal astrometric errors of $0\farcs55$. However, we note that mm1, and also the source c3mm-s11 detected at 3-mm wavelength by \citet{2002ApJ...566..945B}, were observed at lower resolution than 13-s. Each of mm1 and c3mm-s11 cover at least $\sim$5 arcsec, including contributions from other sources outside the region containing the masers reported here.  If the source associated with the Eastern methanol masers is at an early stage of evolution, this may explain why is too faint to be detected as a distinct 1.2-mm source between the Eastern red- and blue-shifted methanol features.

If the water masers we detect toward the IRAS 19410+2336 region were  associated with a bipolar outflow from the same origin, s-13, the jets must be precessing or have an opening angle of $\sim$85$^{\circ}$, and the masers seen at different epochs are in most cases either different ejecta, or interactions with different parts of the surrounding medium. Thus, the outflow could be characterized as poorly collimated. \citet{2007prpl.conf..245A} found that an opening angle of $>50$$^{\circ}$ is usually associated with more evolved sources than those producing well-collimated outflows. {It is possible, therefore, that the water masers are associated with an outflow from 13-s, poorly collimated due to age or the competative accretion mechanism, but this is not consistent with other observations of this region and we consider the alternative possibility that they instead trace a disc. }

\subsection {Disc and infall}
\label{sec:discandinfall}
The presence of circumstellar gas discs around young stars was inferred to be very common by \citet{1993prpl.conf..521B}. The 6.7 GHz methanol maser commonly traces the circumstellar gas disc or outflow of high-mass star-forming regions \citep[and references therein]{2014PASJ...66...31F}.
Observations by \citet{1998ApJ...508..275N} added support to the hypothesis that the 6.7 GHz methanol masers lie within rotating edge-on circumstellar discs.
\par The combined distribution of the methanol masers we mapped in
this region of IRAS 19410+2336 does not appear to be a disc, as the
most red-shifted emission has blue-shifted emission both to the East and
to the West. As explained in Sec. 4.1, the Eastern masers may trace an
  outflow from mm1. \citet{2009A&A...502..155B} found that at high
  resolution, 6.7 GHz methanol masers trace a variety of morphologies
  including infall, and it is possible that the Western masers close
  to 13-s may be a sign of infall.

The closest methanol and water
masers are clumps 14 and d, separated by
96 mas, equivalent to 205 au, 0.001 pc.  Their velocities only are 0.17 km s$^{-1}$ apart. Although this is much closer than the
separations measured at lower resolution by
\citet{2018MNRAS.474.3898B} and \citet{2002A&A...390..289B}, we note
that, as discussed by \citet{2002A&A...390..289B}, the different
pumping mechanisms mean that they are likely to trace different protostellar phenomena \citep[e.g][]{2012A&A...541A..72B,2002A&A...390..289B,2010A&A...517A..71S}, although these can be associated with the same exciting source. The methanol masers are separated from the water masers by more than half the total size of the water maser area, the combined astrometric error is much smaller (16 mas) and the physical separation may be greater than 205 au if the masing clumps are at different distances along the line of sight.  Nonetheless, the masers have appeared at different locations at different epochs. Given the variability of the masers and the excitation of emission at different locations at different epochs, as well as proper motions, it is possible that methanol and
water masers could appear closer in future observations.
\par
{Section~\ref{sec:outflow} explains that, whilst it is possible that the water masers trace an outflow, this is not consistent with other studies which suggest that an outflow from the central, high-mass YSO would be more tightly collimated.}
Although the 22-GHz water maser is not widely known to trace discs or rotating material around protostellar objects, this is not unknown, e.g. \citet{2006PASJ...58..883I} found  strong evidence for water masers tracing rotating material (i.e. a disc) through monitoring the high-mass star forming region G192.16-3.84.
Figure \ref{fig:water_all} illustrates all published H$_2$O masers imaged towards the IRAS 19410+2336 region, which lead us to suggest that they are tracing  rotating gas. At all epochs, these masers
show a consistent offset between more blue-shifted emission to the North and red-shifted to the South (Figure~\ref{fig:water_all}). The maximum angular extent is equivalent to $\sim$350 au
at 2.16 kpc, and the mid point is close to the mm source 13-s \citep{2012A&A...545A..51R}, within its
astrometric uncertainty. If the position change of the MERLIN -- e-MERLIN maser features is due to physical rotation, this corresponds to $V_{\mathrm{pm}}$  =  36 km s$^{-1}$ in a tilted disc of approximately 175 au radius. The $V_{\mathrm{lsr}}$ velocities span $\sim$15 km $^{-1}$, or a line-of-sight velocity of magnitude 7.5 km s$^{-1}$ with respect to the mid-point, giving a total orbital velocity $\sim$37 km $^{-1}$. However, this would correspond to an enclosed mass $\sim$270 M$_{\odot}$,
which is far greater than the mass estimate for 13-s of 8.1 M$_{\odot}$ by \citet{2012A&A...545A..51R}.
Moreover, although the \citet{2002A&A...390..289B} maser positions have large astrometric errors, the
position angle between the red- and blue-shifted emission has a smaller uncertainty, and is inconsistent with rotation in a consistent direction. This makes it more likely that the water maser position changes are due to different parts of the disc being excited at the different epochs.
\par Thus, if we assume that the maser positions observed in 2005 represent the maximum extent of the disc, and deduce from the distribution of masers at all epochs that the angle of inclination with respect to the plane of the sky is $\sim60^{\circ}$, then the rotation velocity is $\sim$15 km s$^{-1}$. This corresponds to a more feasible enclosed mass of $\sim$44 M$_{\odot}$. The main uncertainties are due to the extent and nature of the disc, which could be warped or flared, and the possibility of infall or outflow as well as rotation. If the disc is close to edge-on and the apparent E--W extensions are due to flaring or ablation, the enclosed mass could be as low as 11 M$_{\odot}$, close to the estimate by  \citet{2012A&A...545A..51R}.

\section{Conclusions}
Methanol class II 6.7-GHz and water 22-GHz masers were observed toward IRAS 19410+2336 region at high angular resolution using
MERLIN and e-MERLIN. {The astrometric uncertainty is 12--10 mas, the most accurate measurements yet made in this region.}

The CH$_3$OH maser features span a velocity range of about 16 to 27 km s$^{-1}$ with a peak flux density $\sim$50 Jy. They are spread over a region $\sim1\farcs3$ corresponding to $\sim$2808 au at a distance 2.16 kpc (Figure \ref{fig:Meth_distribution}). They are separated into two groups $0\farcs8$ apart, which might  indicate that our methanol masers  are associated with more than one protostar in this region. {The Eastern group of CH$_3$OH masers have an elongated distribution with a red-blue offset similar to the direction of `C' (see arrows in Figure \ref{fig:all_features}) proposed previously by \citet{2003A&A...408..601B}.  The apparent centre of the Eastern methanol maser outflow is at about $\alpha$(J2000)  = 19$^h$ 43$^m$ 11.26$^s$  $\delta$(J2000) =  +23$\degree$ 44$^\prime$ $03\farcs3$ which is within the mm1 peak position uncertainty of $\sim0\farcs5$. }

\par
The H$_2$O masers, observed in 2005 and 2019, span a velocity range from 13 to $\sim$29 km s$^{-1}$. The peak flux density was found to be 18.7 Jy and 13.487 in 2005, and 2019, respectively. They are distributed over 165 mas ($\sim$350 au at 2.16 kpc). The  H$_2$O masers may be associated with rotating material (a tilted disc) around a protostellar object centered at $\alpha$(J2000) = 19$^h$ 43$^m$ 11.184$^s$  $\delta$(J2000) =  +23$\degree$ 44$^\prime$ $2\farcs96$ which coincides with the mm source  13-s within the combined position accuracy of $\sim$$0.''2$ \citep{2012A&A...545A..51R}.
The estimated mass for this protostar is  11--44 M$_{\odot}$, the lower limit being close to the estimate by \citet{2012A&A...545A..51R}.
The Western methanol masers are all slightly blue-shifted,  and if they are also associated with 13-s it is possible that they are in infall.

This scenario is consistent with the multiple outflows identified by \citet{2003A&A...408..601B} in the IRAS 19410+2336 region, and with the development of multiple protostars. Our findings also imply that the Eastern masers, where no H$_2$O is detected, are associated with a less evolved protostellar object than the Western masers. The angular separation between the two putative protostellar objects is $\sim$$0.''36$, while the physical separation at 2.61 kpc would be at least $\sim$780 au.

\par Eventually, high angular resolution observations of other maser
species in the region are needed to investigate which phenomena they
trace, their association with the embedded protostars and provide
clues to their evolutionary stages in the massive star forming region
IRAS 19410+2336. Observations of dust and thermal lines at higher resolution, ideally allowing a position and astrometric accuracy of 100 mas or better, are needed to  confirm the associations between discs or outflows and the protostars (as in the ALMA observations by \citet{2019A&A...623A..77S} of G023.01-00.41).  ALMA -- or NOEMA, given the Declination of IRAS 19410+2336 -- could make this possible.

\section*{acknowledgements}
We warmly thank Prof. R. Battye, M. Gray and  R. Beswick, and the rest  of the e-MERLIN team for guidance in reducing these data. We also remember the important role of the late Dr Jim Cohen in the initiation of this project. We thank the referee for very helpful comments which have improved this paper. e-MERLIN is the UK radio interferometer array, operated by the University of Manchester on behalf of STFC.
We acknowledge the use of MERLIN archival data as well as  NASA's Astrophysics Data System Service.
M.Darwish would like to acknowledge the Science and Technology Development Fund (STDF) N5217, Academy of Scientific Research and Technology (ASRT), Cairo, Egypt and Kottamia Center of Scientific Excellence for Astronomy and Space Sciences (KCSEASSc), National Research Institute of Astronomy and Geophysics. Our sincere thanks to H. Beuther and F. Widmann for their helpful discussion.

%
\bibliographystyle{mnras}
\bibliography{ref}
%
 \bsp	
\label{lastpage}
   \end{document}